\documentclass[conference]{IEEEtran}
\usepackage{cite}
\usepackage[pdftex]{graphicx}
\usepackage{amsmath}
\usepackage{multirow}
\usepackage{algorithm}
\usepackage{algpseudocode}
\usepackage{xcolor}
\usepackage{subfig}
\usepackage{soul}

\usepackage{url}
\begin{document}
%
\title{Scaling and assigning resources on \\ ion trap QCCD  architectures}


\author{\IEEEauthorblockN{Anabel Ovide, Daniele Cuomo and Carmen G. Almudever}

\IEEEauthorblockA{\textit{Computer Engineering Department, Universitat Politècnica de València, Valencia, Spain}}

\IEEEauthorblockA{\textit{Email: aovigon@upv.es,  cuomo.daniele@outlook.com, cargara2@disca.upv.es}}}
\maketitle

\begin{abstract}
Ion trap technologies have earned significant attention as potential candidates for quantum information processing due to their long decoherence times and precise manipulation of individual qubits, distinguishing them from other candidates in the field of quantum technologies. However, scalability remains a challenge, as introducing additional qubits into a trap increases noise and heating effects, consequently decreasing operational fidelity. Trapped-ion Quantum Charge-Coupled Device (QCCD) architectures have addressed this limitation by interconnecting multiple traps and employing ion shuttling mechanisms to transfer ions among traps. 
This new architectural design requires the development of novel compilation techniques for quantum algorithms, which efficiently allocate and route qubits, and schedule operations. The aim of a compiler is to minimize ion movements and, therefore, reduce the execution time of the circuit to achieve a higher fidelity.

In this paper, we propose a novel approach for initial qubit placement, demonstrating enhancements of up to 50\% compared to prior methods. Furthermore, we conduct a scalability analysis on two distinct QCCD topologies: a 1D-linear array and a ring structure. Additionally, we evaluate the impact of the \textit{excess capacity} -- i.e. the number of free spaces within a trap -- on the algorithm performance. 
\end{abstract}
\begin{IEEEkeywords}
scalability quantum computing systems, modular quantum computers, mapping of quantum algorithms, ion trap QCCD. 
\end{IEEEkeywords}
\section{Introduction}

It is widely known that, in theory, quantum computers have the potential to solve intractable problems for classical computers due to quantum mechanics phenomena such as superposition, entanglement and interference. Nowadays, there are various quantum processors based on different qubit implementations, all facing issues such as high error rates, reduced qubit counts, and relatively short coherence times, which limit their computational power. These technologies, known as Noise-Intermediate-Scale Quantum (NISQ) devices, include superconducting circuits~\cite{Huang2020}, 
quantum dots~\cite{PhysRevA.57.120}, neutral atoms~\cite{doi:10.1080/09500340008244052}, photonic qubits~\cite{RevModPhys.79.135}, and ion traps~\cite{HAFFNER2008155}, among others.  

Although currently there is not an outperforming quantum hardware, ion trap platforms offer several advantages such as long coherence times, identical qubits, all-to-all qubit connectivity (in a single trap), and precise qubit control, being a promising candidate not only for quantum computation but also for quantum networking~\cite{10.1117/12.2657155}. However, one of the main challenges, which is shared with all other quantum technologies, is \textit{scalability}. The number of qubits needs to be substantially increased to build fault-tolerant systems and achieve the full computational power that quantum computers can offer. 

Ion traps utilize the energy states of atomic ions such as Ca+ or Yb+ to encode qubits, which are then manipulated through quantum gates facilitated by lasers. These qubits are confined or trapped, typically in a linear arrangement, within specific regions called traps via electromagnetic fields~\cite{HAFFNER2008155}. Current ion trap chips integrate tens of qubits on a single-core architecture, such as the IonQ Forte quantum processor consisting of 36 qubits~\cite{ionq_forte}. However, to develop larger devices, one cannot keep on increasing the qubit counts within a single trap as the addition of qubits highly amplifies the ion chain's vibrational mode, leading to increased heating and diminished overall fidelity during execution.

\begin{figure}[!h] 
     \centering
    \subfloat[]{
        \includegraphics[width=0.72\columnwidth]{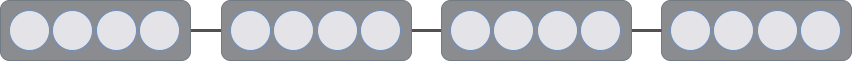}\label{fig:LQCCD}
        }
    \\\vspace{-5pt}
    \subfloat[]{
           \includegraphics[width=0.40\columnwidth]{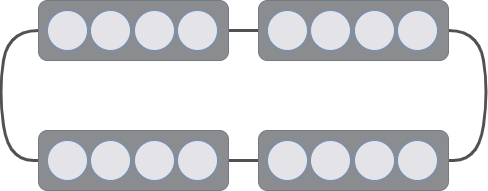}\label{fig:RQCCD}
    }
    \vspace{-7pt}
    \caption{\small{Illustration of (a) a linear and (b) a ring QCCD topologies consisting of 4 traps and 4 ions per trap }}
    \label{fig:Topology}
\end{figure}

Given the constraints in confining more and more qubits within a single trap, Quantum Charge-Couple Devices (QCCD)~\cite{Pino_2021} are a promising approach to scale up ion trap platforms. QCCD interconnects multiple traps using electric fields to shuttle ions between distinct trapping regions. This architectural modular approach allows an increase in the total number of qubits in the device by expanding the number of traps and still maintaining a suitable number of ions per trap. Practical implementations of trapped-ion QCCD-based processors already exist~\cite{PhysRevX.13.041052}, with notable advancements demonstrated by Quantinuum in enhancing scalability\cite{quantinuumScalability}.

The emergence of this new architectural design highlights the need to develop novel quantum circuit mapping techniques. Note that they play a crucial role in optimizing the overall circuit execution. More precisely, they increase the circuit fidelity by reducing the total number of operations and parallelizing them while minimizing the frequency of ion SWAPs (relocation of qubits within a trap) and shuttles (movement of qubits between traps). Notably, mapping methods used in single ion trap devices may no longer be applicable to this multi-trap architecture due to the inherent all-to-all connectivity and lack of shuttle requirements. A few circuit mapping techniques for this new quantum technology have been proposed~\cite{Saki_2022,upadhyay2022shuttleefficient,Schmale_2022,schoenberger2024shuttling,schoenberger2023using}, which involves algorithms for qubit allocation, routing of qubits and operation scheduling.

In this work, we focus on the qubit assignment problem for QCCD architectures and present a novel qubit allocation algorithm, which enhances previous state-of-the-art approaches~\cite{upadhyay2022shuttleefficient,murali2020architecting} by showing an improvement of up to 50\% in the total circuit execution time. This advancement is achieved by considering not only the overall interaction of qubits in the circuit but also the temporal distribution of these interactions. To further explore the potential of this scalable design, we conduct a scalability analysis on two distinct QCCD topologies: a 1D-linear array and a ring structure (Figure~\ref{fig:Topology}). Additionally, we evaluate the algorithm performance based on the impact of the excess capacity, namely the number of free spaces within a trap, set at the beginning of the execution. 

The paper is structured as follows. Section II provides an overview of single and QCCD ion trap devices. In Section III, several mapping techniques for quantum algorithms are presented, with a focus on ion trap processors and highlighting the differences between single and QCCD approaches. Next, in Section IV, our qubit allocation strategy is introduced. In Section V, the selected benchmarks and experimental framework are described. The different experiments and results are presented in Section VI, including a comparison of our qubit allocation method with previously proposed strategies, and its evaluation across different processor topologies. This is followed by a scalability analysis and a study of the excess capacity. The paper ends with the conclusions and future work.
\section{Trapped-ion quantum processors}
Nowadays, numerous quantum technologies are competing to demonstrate the full computational capabilities inherent in quantum information processing. Among these quantum computing platforms, ion trap devices emerge as a promising approach due to their long decoherence times and precise qubit manipulation. In ion trap processors, ions are retained within traps, which are specific regions of the quantum chips designed to confine them via electric and magnetic fields (Penning traps) or an oscillating electric field (Paul traps)~\cite{Bruzewicz_2019}. Typically, ions within traps are arranged linearly. To manipulate the states of qubits and perform operations, including single and two-qubit gates, various techniques can be employed depending on the qubit type, such as optical methods (laser manipulation) or microwave-based approaches~\cite{Bruzewicz_2019}.

The prolonged coherence times offered by ion traps make this technology a promising candidate not only for quantum computation~\cite{PhysRevLett.74.4091, Debnath_2016} but also for quantum network applications~\cite{10.1145/3386367.3431293,doi:10.1126/science.aam9288} including distributed quantum computing ~\cite{Cuomo_2020,10214316}. They entail the execution of quantum programs or algorithms across multiple quantum devices interconnected via quantum and classical links. These executions need constant communications, both classical and quantum, among different processors, which may incur prolonged waiting times as devices must await responses from others to proceed with program execution. These waiting times can potentially lead to qubit state degradation highlighting the suitability of ion trap devices for quantum networking~\cite{toh2023progress,siverns2017ion,PhysRevA.76.062323}. 

Multiple trapped-ion processors have been developed, integrating tens of qubits within a single trap~\cite{ionq_forte,Figgatt2019, holz20202d}. However, scalability poses a significant challenge for this technology. As previously mentioned, qubits are confined via electric and magnetic fields, and increasing their counts makes the confinement process more complex. Moreover, the precision required to perform operations becomes more intricate with the increase in the ions' vibrational states~\cite{murali2020architecting,Bruzewicz_2019}, which also leads to a growth in the system heating. Consequently, as the number of qubits per trap gets larger and hardware complexity rises, the fidelity of executions drastically reduces.

An alternative and promising approach to address this problem is the modular trapped-ion Quantum Charge-Coupled Device (QCCD) architecture consisting of several traps\cite{Pino_2021,PhysRevX.13.041052}. QCCD employs shuttles to connect different traps, enabling ions to be physically transported from one trap to another with high fidelity~\cite{Akhtar2023}. By interconnecting traps in this manner, the total number of qubits can be increased without a significant decrease in fidelity, as a reduced number of ions are confined in each trap. These devices support a variety of potential topologies, including a 1D linear array, ring devices, grid structures~\cite{murali2020architecting}, or X-junction configurations~\cite{Schmale_2022}. Note that this approach is not exclusive to ion trap processors. Next generations of superconducting quantum devices are expected to be multi-core in which multiple quantum chips are interconnected via quantum and classical links~\cite{10.1145/3457388.3458674,9923784,bravyi2022future}.

QCCD ion traps introduce a set of new challenges, encompassing both hardware and quantum circuit mapping methods. Hardware-related difficulties arise from the complexity of the structures and connections between traps. Furthermore, existing quantum algorithm mapping techniques applicable for single-ion trap devices are no longer compatible with this new approach, which will be discussed in the subsequent section.
\section{Mapping of quantum circuits on NISQ devices}
Hardware-agnostic quantum algorithms or programs cannot be directly executed on quantum processors as they do not consider hardware restrictions. The main limitations arise from gates, which necessitate decomposition into the ones supported by the quantum chip, and from the restricted qubit connectivity, which limits possible interactions requiring qubit relocation for executing two-qubit gates. To overcome these constraints, quantum circuit mapping techniques are employed to translate and accommodate the quantum circuit to the designated quantum hardware. They usually consider multiple quantum processor characteristics,  including constraints related to qubit connectivity, gate timing, parallel execution of operations, idle time minimization, resource availability, gate fidelity, and qubit coherence times. 



Note that the quantum circuit mapping process is crucial for maximizing the algorithm performance, given the impairments of current quantum devices.

\subsection{The quantum circuit mapping process}

The mapping of quantum circuits or algorithms has three fundamental steps across all quantum technologies: qubit allocation or initial placement, qubit routing, and operations scheduling.

\textbf{Allocation of qubits or initial qubit placement}. In this step, logical qubits (i.e. qubits in the quantum circuit) are strategically assigned to physical qubits in order to reduce and facilitate subsequent movements during routing. This paper will focus on this stage and introduce a novel initial qubit placement algorithm that will be explained in subsequent sections. 

\textbf{Routing of qubits}. Given the qubit connectivity constraint, qubits may need to be relocated within the quantum chip to perform two-qubit gates. For instance, in superconducting devices, qubits must be adjacent to perform two-qubit gates,  whereas in trapped-ion QCCD architectures, qubits involved in a two-qubit operation must be placed within the same trap. In this process, additional gates for `moving' qubits around are introduced to relocate ions. The number of movements should be minimal, as they result in a longer execution time and an increase in total operations, consequently decreasing the overall execution fidelity. 

\textbf{Scheduling of the operations}. This step aims at minimizing the execution time of the quantum circuit by maximally parallelizing quantum operations. It typically involves the utilization of a quantum operation dependency gate graph (nodes represent gates and edges dependencies between them). 

Considering that different quantum technologies are subject to different constraints — for instance, the requirement in superconducting devices for qubit adjacency when executing two-qubit gates and the insertion of the corresponding SWAPs or the necessity of shuttling ions in  QCCD architectures — the quantum circuit mapping procedure should be adapted accordingly.

\subsection{From single-trap to  multi-trap architectures}

An important feature of ion trap technology is the presence of all-to-all connectivity within each trap. This property simplifies the mapping process in single ion trap devices, as logical qubits can be randomly assigned to physical qubits (qubits remain stationary during operations). Furthermore, qubit routing is unnecessary, as operations can be executed without the need for ion rearrangement. Therefore, the scheduling process becomes the primary focus during mapping, as it involves the strategic prioritization of available gates for execution. 

\begin{figure}[!h] 
\centering
    \subfloat[]{
        \includegraphics[width=0.25\columnwidth]{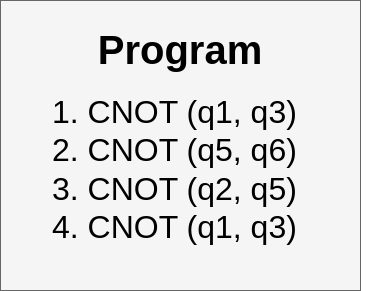}\label{fig:ex_struc}
        }
    \subfloat[]{
        \includegraphics[width=.38\columnwidth]{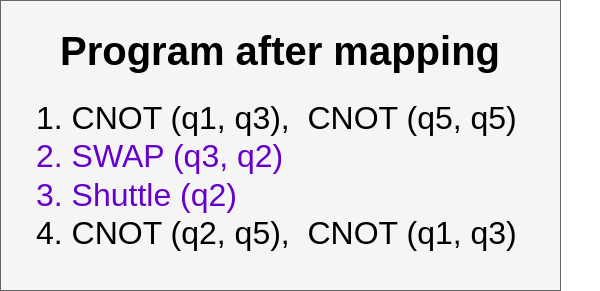}\label{fig:ex_ex}
    }
    \subfloat[]{
           \includegraphics[width=0.35\columnwidth]{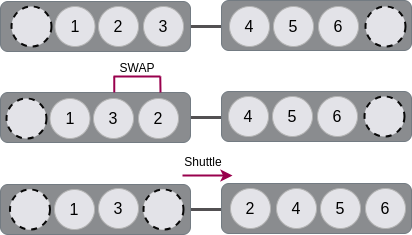}\label{(fig:ex_or)}
    }
    \vspace{-5pt}
    \caption{\small{Example of mapping a quantum circuit to a 1D-linear topology composed of two traps with a capacity of 4 ions each. (a) Program to be executed on the device. (b) Program after the mapping procedure in which two operations, a SWAP and a SHUTTLE (in purple), have been inserted. (c) Overview of the mapping process: (i) qubit allocation in which the first two CNOTs can be directly performed. Note that they can be performed in parallel as the corresponding pair of qubits are allocated in distinct traps; (ii) qubit 2 needs to be moved to the trap on the right to continue execution. It requires adding a SWAP gate between qubits 2 and 3 to position the ion at the end of the trap to be (iii) shuttle to the adjacent trap. After these extra operations, the last two CNOT gates can be performed.}}
    \label{fig:Mappingexample}
\end{figure}

The complexity of the mapping procedure notably scales when considering ion trap QCCD architectures. In the initial qubit placement stage, qubits have to be optimally placed to minimize or even eliminate the need for future shuttling between traps. Moving qubits from trap to trap will require not only adding shuttling operations but also qubit relocation within the trap itself, as the ions to be shuttled need to be positioned in the trap extreme for being transferred. The routing of qubits also becomes complex,  as it must account for both shuttles required to transport ions throughout the device and for the optimal placement of qubits within the trap. In addition, the scheduler must now contemplate the additional gates introduced to facilitate ion reallocation within a trap and the newly added shuttling movements. Note that in current QCCD architectures, parallel operations within a single trap are not allowed. However, parallel execution of non-dependent operations is possible when these are performed in qubits allocated in different traps. It is worth noting that the maximum number of parallel operations corresponds to the number of traps the device has. An example of mapping a quantum circuit on a QCCD architecture is shown in Figure~\ref{fig:Mappingexample}.

Currently, a few mapping methodologies have been proposed for efficiently executing quantum circuits on QCCD  architectures. For instance, Saki et al.~\cite{Saki_2022} proposed a scheduling and qubit routing approach, which are based on the number
of gates a qubit has in different traps, aiming to identify the most efficient ion movement between traps; it also considers full traps when moving ions, trying to avoid or decongest them. Upadhyay et al.~\cite{upadhyay2022shuttleefficient} introduced an initial qubit placement that considers future operations when placing the qubits. Additionally, in~\cite{Schmale_2022}, the authors propose a circuit mapping procedure for a four-trap architecture interconnected by an X junction. Furthermore, a routing algorithm formulated as a Boolean satisfiability (SAT) problem was proposed by Daniel et al. in~\cite{schoenberger2023using}, improving their method in subsequent work~\cite{schoenberger2024shuttling}.

Further research of the mapping procedure for scalable trapped-ion architectures is needed not only to efficiently execute quantum algorithms but also, and more importantly, to investigate the potential of new modular architectural designs and guide future developments. In this paper, we introduce a novel initial qubit placement that improves previous qubit allocation strategies by considering the number of qubits interacting with each qubit, the frequency of these interactions, and when those interactions take place (time domain), as will be explained in the next section. In addition, a scalability analysis is conducted in which weak and strong scaling experiments are performed. Furthermore, the initial optimal excess capacity is studied.
\section{Qubit allocation for QCCD architectures}
State-of-the-art initial qubit placement techniques for ion trap QCCD processors have been proposed by Murali et al.~\cite{murali2020architecting} and Upadhyay et al.~\cite{upadhyay2022shuttleefficient}. These methodologies rely on a qubit interaction graph (see Figure~\ref{fig:STAexample}b), where nodes denote qubits and edges represent interactions between them, describing two-qubit gates. The edges are weighted, indicating the number of two-qubit gates between each qubit pair. In both works, logical qubits are assigned to physical traps based on the edge weight, placing edges with higher weights first. 


The main distinction between these two approaches resides in the use of gate prioritization. More precisely, Upadhyay's method prioritizes gates occurring at the initial stages of the program over those appearing later, updating the weight of the edges based on this information. This is achieved by utilizing a decaying function for any re-occurrence of gates, assigning a negative value after a certain number of operations. However, they both overlook the consideration of the total number of qubits interacting with each qubit, which serves as an indicator of the extent to which a qubit will need to be relocated. Specifically, for a given qubit, a higher and more diverse number of operations with other qubits will potentially result in an increase in movements to reach the traps where other qubits are placed, compared to qubits with fewer interactions. In addition, Upadhyay's method, while considering the timing of operations (i.e. when the qubit interactions take place), is suboptimal for lengthy algorithms due to the introduction of a negative weight for late occurrences. Moreover, they do not incorporate ion repositioning within traps based on the operations they are involved in and the temporal distribution of those operations, thereby minimizing ion movements within traps for potential shuttling operations.

In this work, we introduce a novel initial qubit placement strategy, the \textit{``Spatio-Temporal Aware Qubit Allocation Algorithm"} (STA). STA enhances previous qubit allocation approaches by incorporating not only the number of gates between qubit pairs and the temporal aspect of interactions but also the ratio of qubit-to-qubit interactions. Additionally, it includes a reordering of qubits within traps considering the time operations are executed.

\subsection{The Spatio-Temporal Aware Qubit Allocation Algorithm}
The Spatio-Temporal Aware Qubit Allocation Algorithm (STA) contemplates: (i) the involvement of qubit pairs in two-qubit gates while (ii) considering the time in the circuit in which the operations take place, and (iii) the ratio of qubit interactions for each qubit in the circuit. It also considers the initial excess capacity of each trap, which is defined as the number of free spaces within a trap~\cite{Saki_2022}, at the beginning of the execution (i.e. in the routing of the qubits, those free spaces can be used). The pseudocode of the STA algorithm is shown in Algorithms~\ref{alg:STA} (main routine),~\ref{alg:mapQubit}, and~\ref{alg:orderQubits}.

The algorithm works based on two data structures: the \textit{qubit interaction ratio (R)}, consisting of $N$ elements (being $N$ the number of logical qubits) and the \textit{temporal weight (T)}, whose values are calculated for each interacting pair of qubits considering the gate execution time. The two lists are derived as follows. After obtaining the operations per time slice (Algorithm~\ref{alg:STA} line 1, \textit{``S"}), which are the operations that can be performed at a specific time in the circuit execution, the ratio of qubit interactions $R(q_i)$ for each qubit and the temporal weight $T(q_i,q_j)$ for each interacting pair of qubits can be computed. The qubit interaction ratio (Algorithm~\ref{alg:STA} line 2, \textit{"R"}) is calculated as:
\begin{equation}
    R(q_i)=\frac{r_i}{N}
\end{equation}
where $r_i$ represents the number of different qubits $q_i$ interacts with, and $N$ the number of logical qubits involved in the program. This calculation will give information about how much each qubit interacts with others, a crucial consideration when positioning qubits within traps. Qubits with higher interaction ratios will necessitate more movements, thus optimizing their placement becomes crucial.

Subsequently, a value for each interacting pair of qubits (Algorithm~\ref{alg:STA} line 3, \textit{ ``T"}) is calculated. This value is based on the number of two-qubit gates involving a pair of qubits within the respective time slice. The temporal aspect is determined by the following formula, which is inspired by the work of Baker et al.~\cite{Baker_2020}:

\begin{equation}
    T(q_i,q_j)=\sum^sI(s,q_i,q_j)\times2^{-s}
\end{equation}

where $q_i$,$q_j$ denote the interacting pair of qubits, $s$ represents the current time slice (starting from 0), and the function $I(s,q_i,q_j)$ is equal to zero if the specified qubits do not interact during the given time slice; otherwise, it is equal to one. $2^{-s}$ is the lookahead component (considers time). A duplicate of this list is generated for the purpose of relocating qubits within traps at the end of the algorithm ($T_{copy}$).

\begin{algorithm} [b]
\caption{STA}\label{alg:STA}
    \hspace*{\algorithmicindent} \textbf{Input: } $C$ \Comment{Quantum circuit\ \ \ \ \ \ \ }
    \begin{algorithmic}[1]
        \Statex \textit{Global lists}
        \State $S \gets \text{computed from } C$ \Comment{Slices\ \ \ \ \ \ \ \ \ \ \ \ \ \ \ \ \ \ \ \  } 
        \State $R \gets \text{computed from } S$ \Comment{Ratio info\ \ \ \ \ \ \ \ \ \ \ \ \ \ \ }
        \State $T \gets \text{computed from } S$ \Comment{Spatio-Temporal info}
        \State $T_{\text{copy}} \gets  T  \text{ in reverse order}$
        \While{\textit{R}}
            \State $\text{\textbf{map\_qubit}}(R[0])$\Comment{Qubit with highest ratio}
        \EndWhile
        \State $\text{\textbf{order\_qubits}}(T_{\text{copy}})$
\end{algorithmic}
\end{algorithm}

\begin{algorithm}
\caption{map\_qubit}\label{alg:mapQubit}
    \hspace*{\algorithmicindent} \textbf{Input:} $q_1$ 
    \begin{algorithmic}[1]
        \For{$q_{pair} \in \textit{T}$}
            \If{$q_1  \in q_{pair}$}
                \State $q_2 \gets (q_1 \neq q_{pair}[0]) \text{ ? } q_{pair}[0] : q_{pair}[1]$
                \State Break
            \EndIf
        \EndFor
        \If{$q_2 \in T [:\text{pos}(q_{pair})]$} \Comment{Check if $q_2$ before in list}
            \State \textbf{map\_qubit($q_2$)}
        \EndIf
        \If{$ q_1 \in R \text{ or } q_2 \in R$}
            \State place\_shortest\_path($q_1$,$q_2$)
            \State $R.\text{remove}(q_1,q_2)$ \Comment{Removes qubit/s placed}
            \State $T.\text{remove}(q_1,q_2)$ \Comment{Removes pair\ \ \ \ \ \ \ \ \ \ \ \ \ \ }
        \EndIf
        \State Return
\end{algorithmic}
\end{algorithm}

\begin{algorithm}
\caption{order\_qubits}\label{alg:orderQubits}
    \hspace*{\algorithmicindent} \textbf{Input:} $T_{\text{copy}}$
    \begin{algorithmic}[1]
        \For{$q_{pair} \in  T_{\text{copy}}$}
            \If{$q_{pair} \in \text{trap}$}
                \State Continue
            \Else
                \State place\_shortest\_path($q_1$,$q_2$) \Comment{Within their trap}
            \EndIf
        \EndFor
\end{algorithmic}
\end{algorithm}

\begin{figure}[!h] 
\centering
    \subfloat[]{
         \includegraphics[width=0.6\columnwidth]{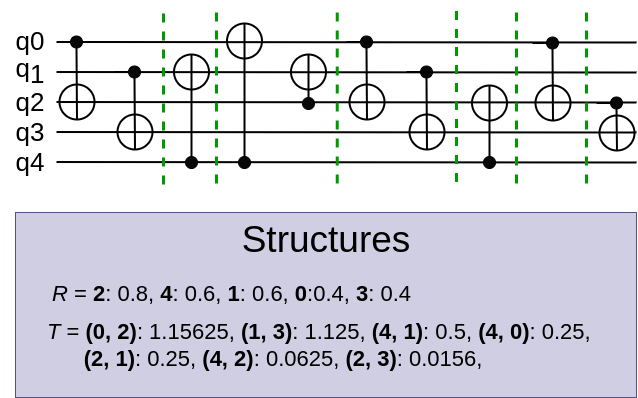}\label{fig:ex_struc}
    }
    \subfloat[]{
        \includegraphics[width=0.35\columnwidth]{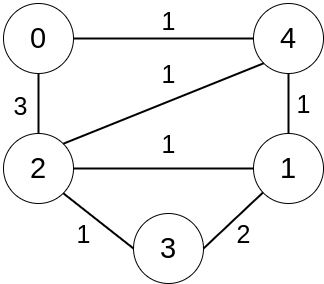}\label{fig:ex_ex}
    }
    \vspace{-5pt}
    \caption{(a) Quantum circuit to be executed along with the initial structures; green dash lines illustrate the different time slices. The "R" structure contains the qubit ratio values, while the "T" structure holds information about the qubit pair interaction considering the time aspect. (b) Qubit interaction graph.}
    \label{fig:STAexampleInit}
\end{figure}

\begin{figure*}[t!] 
\centering
    \subfloat[]{
        \includegraphics[width=1.2\columnwidth]{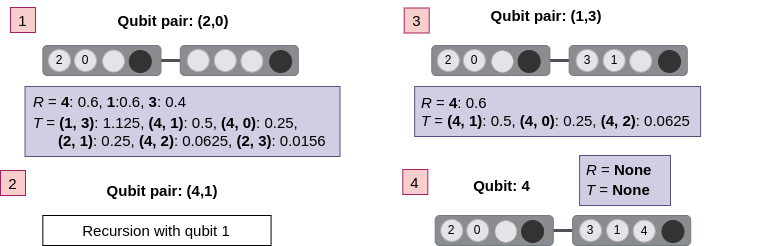}\label{fig:ex_ex}
    }
    \subfloat[]{
           \includegraphics[width=0.45\columnwidth]{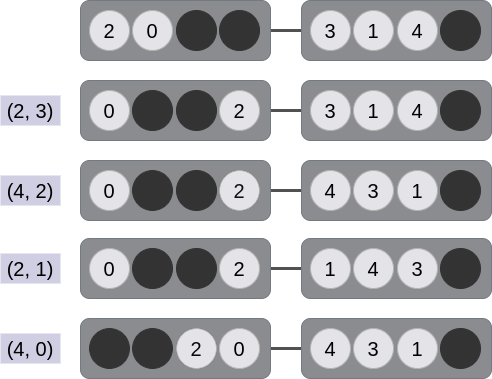}\label{(fig:ex_or)}
    }
    \vspace{-7pt}
    \caption{Example of the STA algorithm for a device consisting of two traps and 4 ions per trap, traps possess an excess capacity of two free spaces. (a) Steps involved in positioning qubits within traps. (b) Qubit relocation process.}
    \label{fig:STAexample}
\end{figure*}

 An illustrative example of how STA allocates qubits on a 2 trap device is shown in Figure~\ref{fig:STAexampleInit} and Figure~\ref{fig:STAexample}. Given a quantum circuit (described as a list of gates), the structures \textit{R} and \textit{T} are derived (Figure~\ref{fig:STAexampleInit}(a)). Observe that qubit 2, in \textit{R} list, demonstrates the highest ratio (4/5) by computing the different qubit interactions (4) divided by the number of qubits (5).In list  \textit{T}, the pair (0,2) is positioned at the beginning of the list, as those qubits are involved in the first gate to be executed and exhibit the highest number of two-qubit gate operations between them.

Once the two data structures are calculated, the algorithm will run until there are no elements remaining in the qubit ratio list (Algorithm~\ref{alg:STA} line 5-7). It operates recursively (Algorithm~\ref{alg:mapQubit}), prioritizing the placement of qubits with higher interaction ratios. Therefore, the first input will be the qubit with the highest ratio (first element in \textit{R}); qubit 2 in the provided example. Followed by subsequent qubits with progressively lower ratios.

 Each time the function \textit{map\_qubit} is called, a second qubit forming a pair with the input qubit is identified from the $T$ structure (Algorithm~\ref{alg:mapQubit} line 3, $q_2$), being qubit 0 the corresponding pair of qubit 2 following the example. Subsequently, it is verified whether this identified qubit has no interactions closer in time or a higher number of operations with another qubit; if it does, the algorithm is recursively called with the qubit that interacts nearest in temporal proximity (algorithm~\ref{alg:mapQubit} line 7-9). Qubit 0 does not appear earlier in \textit{T} list, suggesting that it has not previously engaged in any interaction and/or has participated in a significant number of operations with qubit 2, being both scenarios in this case.

Following the confirmation that the pair of qubits is the closest in time or has the higher number of operations compared to other potential candidates regarding qubit interaction, both qubits are placed within the same trap if none of them have been previously allocated and removed from the lists. Suppose only one of the qubits has already been placed. In that case, the other one is allocated by considering the shortest path between them using the Dijkstra algorithm and afterward removed from both lists (algorithm~\ref{alg:mapQubit} line 10-14). In the provided example, no qubits have been placed yet. Thus, both qubits are assigned to the same trap (Figure~\ref{fig:STAexample} (b)), and their respective entries are removed from the lists. These steps will be repeated until all qubits are allocated. 

Continuing with the example, the subsequent input for the function is qubit 4 in the \textit{R} list. Similar to previous steps, the pair for this qubit is searched in \textit{T}, as performed before, resulting in qubit 1. Afterward, it is verified whether qubit 1 has interacted previously with another qubit, resulting in qubit 1 having a higher value with another qubit (i.e. (1,3)), so the function is called recursively with qubit 1. 
Following this recursive process, it is determined that qubit 1 interacts with qubit 3, which either has no previous interactions or has participated in a significantly limited number of operations with other qubits. Consequently, both qubits, 3 and 1, are placed in the same trap. Since the first trap does not have enough space for both qubits (considering the excess capacity), they are instead placed in the second trap and removed from the lists.
Upon completing the recursion, the focus of the function shifts to qubit 4. With the remaining qubits already placed, qubit 4 is allocated to the second trap, thereby concluding the recursion process. The main algorithm proceeds by invoking the function responsible for reordering the qubits within traps.

Once all qubits are placed (i.e., when there are no more qubits in the \textit{"R"} structure), they will be reordered within each trap (algorithm~\ref{alg:STA} line 8, algorithm~\ref{alg:orderQubits}). This arrangement is accomplished by iterating through the copy of the interaction qubit pairs structure created earlier in a reverse order (algorithm~\ref{alg:STA} line 4, $T_{copy}$). Each qubit pair ordered from the lowest value to the highest one will be checked whether they are placed within the same trap (algorithm~\ref{alg:orderQubits} lines 2-4). If that is not the case, qubits will be relocated to the extremes of the trap closest to the trap the other qubit is (algorithm~\ref{alg:orderQubits} lines 5-6). After the relocation process for all qubits is completed, the STA algorithm will be finalized.

As illustrated in Figure~\ref{fig:STAexample} (b), by iterating through \textit{T} structure, the first pair of qubits is selected (2,3), and it is checked whether they are situated within the same trap. Since this is not the case, both qubits are re-positioned at the end of the trap closer to the other qubit, resulting in the shortest path between them.
Subsequently, the next pair of qubits, (4,2), is considered. Qubit 2 does not require relocation, whereas qubit 4 is now positioned at the end of the trap as done previously with qubit 3. Notably, qubit 3 becomes adjacent to qubit 4, meaning that qubits with closer execution times or more operations are prioritized, followed by subsequent ones. The process continues until all qubits are relocated, if necessary.

The STA algorithm, though presented initially for linear and ring topologies, can be applied to a variety of structures such as those that have X/T junctions. Furthermore, while it was initially designed for ion trap technology, its applicability is not limited solely to these quantum processors. The algorithm can also be employed in other modular quantum computing architectures, such as multicore (multinode) superconducting quantum devices~\cite{10181589,Baker_2020,10313719,10.1145/3655029}, where the initial placement provided by STA serves as an initial strategy for allocating qubits across the different cores.

\section{Simulation framework and benchmarks}
The experiments of this work were conducted using the QCCDSim framework provided by ~\cite{murali2020architecting}, a simulator for ion-trapped QCCD architectures, offering comprehensive features, including information about heating considerations, and execution times and fidelities, among others. This simulator also allows the implementation of custom mapping techniques, as the initial qubit placement proposed in this paper. Moreover, it gives the flexibility to utilize various trapped-ion QCCD topologies and performance metrics.

To perform experiments with the STA algorithm, an already implemented scheduling and qubit routing method, presented in ~\cite{Saki_2022}, has been used. The routing is based on the number of gates a qubit has in different traps. This analysis aims to identify the most efficient movement, determining which ion should be transferred to another trap. 

\begin{table}[b!]
    \centering
    \vspace{-3pt}
    \caption{Benchmarks}
    \label{tab:Benchmarks}
    \vspace{-8pt}
    \begin{tabular}{*{4}{c}}
        \multirow{2}{*}  & \multirow{2}{*}{\textbf{Slices}} & \multirow{2}{*}{\textbf{Two-qubit gates}} & \multirow{2}{*}{\textbf{Av. Gates/Slice}} \\
        & & & \\ [-1ex]
        \hline
        \textbf{CA}  & 451 & 513 & 1.14 \\
        \textbf{DA}  & 185 & 1520 & 8.22 \\
        \textbf{QAOA}  & 125 & 2016 & 16.128 \\
        \textbf{QFT}  & 125 & 2016 & 16.128 \\
        \textbf{QV}  & 192 & 6144 & 32 \\
        \textbf{RND} & 81 & 991 & 12.23 \\
        \hline
    \end{tabular}
\end{table}


The scheduler works in an earliest-ready gate-first approach as previously done in~\cite{murali2020architecting}, where the order is determined by the gate dependency graph. 

A set of diverse algorithms has been chosen to benchmark STA for different scalable QCCD architectures. It includes the Cuccaro Adder (CA), the Draper Adder (DA), the Quantum Approximate Optimization Algorithm (QAOA), the Quantum Fourier Transform (QFT), Quantum Volume (QV), and Randomly generated Circuits (RC) in which experiments were conducted 20 times. The selection of these benchmarks was driven by their structural diversity, as shown in Table~\ref{tab:Benchmarks}. Algorithms like the CA, DA, QAOA, and QFT are more structured. The CA is characterized as the lightest due to its notably fewer two-qubit gates compared to other algorithms, conversely to the QFT and QAOA algorithms, which also exhibit a high average of two-qubit gates per slice. Additionally, their degree of parallelization varies; QAOA and QFT exhibit the highest level of gate parallelization (average number of gates per slice, last column in Table I), whereas CA is a more sequential algorithm.

Conversely, unstructured algorithms, such as RC and QV, incorporate randomness. Among these benchmarks, Quantum Volume stands out as the heaviest, featuring a significant number of two-qubit gates, slices, and the highest gate parallelism.

The execution of the benchmarks was simulated on two distinct topologies: a 1D linear array and a ring structure (Figure~\ref{fig:Topology}). The selection of a linear topology was based on previous works~\cite{murali2020architecting,Saki_2022}; that is, to be able to accurately compare the STA approach with prior initial qubit placement strategies. The ring structure was chosen as a natural extension of the linear topology, offering enhanced trap connectivity. Moreover, it actually represents a configuration utilized in the Quantinuum H2 QCCD system~\cite{QuantinuumH2}, currently the most promising ion-trap technology.

\section{Evaluation and Scalability analysis}
The Spatio-Temporal Aware Qubit Allocation Algorithm (STA) will be first evaluated against the Greedy policy used in~\cite{murali2020architecting}  as well as a Random approach, wherein qubits are randomly assigned to the traps, in a linear architecture. Moreover, comparisons will be conducted between the 1D-linear array and a ring structure using the STA algorithm. In addition, an architectural scalability analysis will be conducted, in which we will also explore different excess capacities. For all experiments, the QCCD simulator~\cite{murali2020architecting} will be utilized, alongside the qubit routing approach proposed in~\cite{Saki_2022}. The principal used metric is the execution time, which includes the duration required to execute all essential operations such as gates, shuttling, merge, and split. Additionally, specific experiments provide the number of inserted SWAPs and performed shuttles. For further details on how these metrics are computed, refer to~\cite{murali2020architecting}.

\subsection{Qubit allocation strategies: Random vs Greedy vs STA }
\begin{table*}[h]
    \centering
    \vspace{-3pt}
    \caption{Initial qubit placement comparison (linear topology)}
    \label{tab:IP_comparison}
     \vspace{-5pt}
    \begin{tabular}{*{13}{c}}
        \cline{2-10}
\multirow{-1}{*}{} & \multicolumn{3}{|c|}{\textbf{Random}} & \multicolumn{3}{c|}{\textbf{Greedy}} & \multicolumn{3}{c|}{\textbf{STA}} \\
        \cline{2-10}
        \multirow{2}{*} & \multirow{2}{*}{\textbf{Shuttles}} & \multirow{2}{*}{\textbf{SWAPs}} & \multirow{2}{*}{\textbf{Time(s)}} & \multirow{2}{*}{\textbf{Shuttles}} & \multirow{2}{*}{\textbf{SWAPs}} & \multirow{2}{*}{\textbf{Time(s)}} & \multirow{2}{*}{\textbf{Shuttles}} & \multirow{2}{*}{\textbf{SWAPs}} & \multirow{2}{*}{\textbf{Time(s)}} & \multirow{2}{*}{\textbf{$\Delta G(\%)$}} & \multirow{2}{*}{\textbf{$\Delta R(\%)$}}\\
        & & & & & & & & & \\ [-1ex]
        \hline
        \textbf{CA} & 168 & 153 & 0.31 & 105 & 69 & 0.2 & 10 & 4 & 0.1 & \textbf{50\%}& \textbf{67.74\%}\\
        \textbf{DA} & 1171 & 1112 & 1.83  & 647 & 616 & 1.48  & 532 & 490 & 0.98 & \textbf{33.78\%}& \textbf{46.45\%}\\
        \textbf{QAOA}  & 3737 & 3605 & 5.34 & 1442 & 1337 & 2.24 & 1016 & 947 & 1.72 & \textbf{23.21\%}& \textbf{67.79\%} \\
        \textbf{QFT}  & 3872 & 3758 & 5.55 & 1442 & 1338 & 2.24 & 761 & 665 & 1.32 & \textbf{41.07\%}& \textbf{76.22\%}  \\
        \textbf{QV}  & 3545 & 3322 & 5.79 & 3336 & 3147 & 5.55 & 3234 & 3104 & 5.48 & 1.26\%& 5.35\% \\
        \textbf{RND} & 1682 & 1555 & 2.33 & 1298 & 1236 & 1.89 & 1323 & 1267 & 1.93 & 2.07\%& 18.88\% \\
        \hline
    \end{tabular}
\end{table*}

The STA algorithm is compared with the Greedy policy outlined in a prior work~\cite{murali2020architecting}, as well as a random approach in which qubits are randomly distributed among various traps. It should be noted that this comparison does not include the qubit allocation method proposed by Upadhyay et al. ~\cite{upadhyay2022shuttleefficient} as its performance is similar to the Greedy approach. 

Building upon prior works~\cite{upadhyay2022shuttleefficient,murali2020architecting,Saki_2022}, an equivalent configuration of qubits and QCCD topology has been selected for comparison. Specifically, all benchmarks will be executed utilizing 64 logical qubits (program qubits) within a 1D-linear array comprising 6 traps, each containing 17 ions. Furthermore, all traps will be equipped with an initial excess capacity of two (i.e. at the beginning of the execution, 15 qubits will be located per trap).

Table~\ref{tab:IP_comparison} shows the outcomes of such comparisons in terms of the number of shuttles required, SWAP operations necessary for ion relocation within the trap, and the total execution time of the circuit derived from the duration of all operations, including shuttling, gates, split and merge. Note that the execution time reveals how fidelity might vary across different approaches, with longer execution times potentially resulting in lower circuit fidelity due to increased operational durations. Furthermore, it offers insights into the potential heat accumulation within the device, as prolonged execution times correlate with heightened heat generation. 

STA improves the overall execution time for all considered circuits compared to the random ($\Delta R$) and the greedy ($\Delta G$) strategies. A notable enhancement is observed compared to random qubit initial placement, reaching up to 76.22\% for the Quantum Fourier Transform (QFT) algorithm. When compared to the Greedy policy, STA exhibits execution time improvements of up to 50\%, as can be seen in the Cuccaro Adder (CA). However, these are less pronounced for the random algorithms.

An important observation is that qubit allocation plays a critical role in light algorithms (few number of two-qubit gates), as the positioning of qubits significantly influences potential movements within the device. Conversely, in heavy algorithms (high number of two-qubit gates), the impact of qubit allocation diminishes. This trend is reflected in our results. For instance, CA exhibits the most significant enhancement as it is the lightest algorithm. Contrarily, heavier algorithms, such as Quantum Volume (QV), possess a notably higher number of operations and a higher average of two-qubit gates per slice, thus diminishing the influence of initial qubit placement.
This logic suggests that reducing the number of qubits or gates of the algorithm could further amplify the benefits of the STA approach. Conversely, as the number of qubits and gates increases, the impact of such improvements becomes less pronounced, a phenomenon applicable to any qubit allocation method. For subsequent experiments, only the STA approach will be employed, given its demonstrated effectiveness.

\subsection{Topological Performance Analysis: Linear vs Ring}

In this section,  a comparison between two different topologies has been performed: a 1D-linear array and a ring. As in the previous experiment, both configurations consist of 6 traps and 17 ions per trap, with an initial excess capacity of two for all traps within the devices.
\begin{table}[h!]
    \centering
    \vspace{-3pt}
    \caption{Ring topology}
    \label{tab:ring_topology}
    \vspace{-8pt}
    \begin{tabular}{*{4}{c}}
        \multirow{2}{*} & \multirow{2}{*}{\textbf{Shuttles}} & \multirow{2}{*}{\textbf{SWAPs}} & \multirow{2}{*}{\textbf{Time(s)}} \\
        & & & \\ [-1ex]
        \hline
        \textbf{CA} & 10 & 4 & 0.1 \\
        \textbf{DA} & 740 & 671 & 1.23\\
        \textbf{QAOA} & 997 & 927 & 1.68 \\
        \textbf{QFT} & 1621 & 1093 & 1.98 \\
        \textbf{QV} & 3092 & 2767 & 5.05 \\
        \textbf{RND} & 2578 & 2306 & 3.57 \\
        \hline
    \end{tabular}
\end{table}

\begin{figure}[b]
    \centering
        \includegraphics[width=0.75\columnwidth]{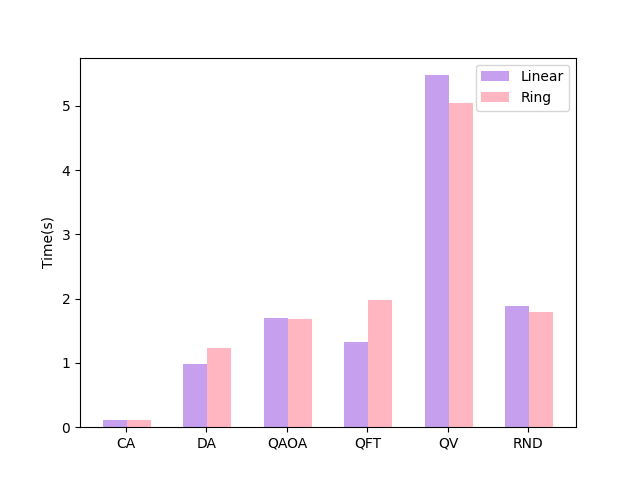}
        \vspace{-10pt}
        \caption{\small{Time execution for a 1D-linear array and a ring topology. Both devices consist of 6 traps, 17 ions per trap, and an initial excess capacity of 2. Benchmarks have 64 logical qubits.}}
        \label{fig:lingVSlinear}
\end{figure}

Table~\ref{tab:ring_topology} presents the metrics obtained for the ring topology. In addition, Figure~\ref{fig:lingVSlinear} shows the overall execution time for both the linear and ring devices. It can be observed that for most benchmarks, except those with a random component, the linear device generally outperforms the ring counterpart. This observation can be attributed to the fact that the routing scheme proposed in~\cite{Saki_2022} has been tailored for linear devices, potentially restricting possible optimizations for ring topologies. This explanation is further supported by subsequent experiments when conducting scalability analyses, where ring devices typically outperform linear ones.

\subsection{Scalability analysis}

\begin{figure*}[!h] 
     \centering
    \subfloat[]{
        \includegraphics[width=0.49\columnwidth]{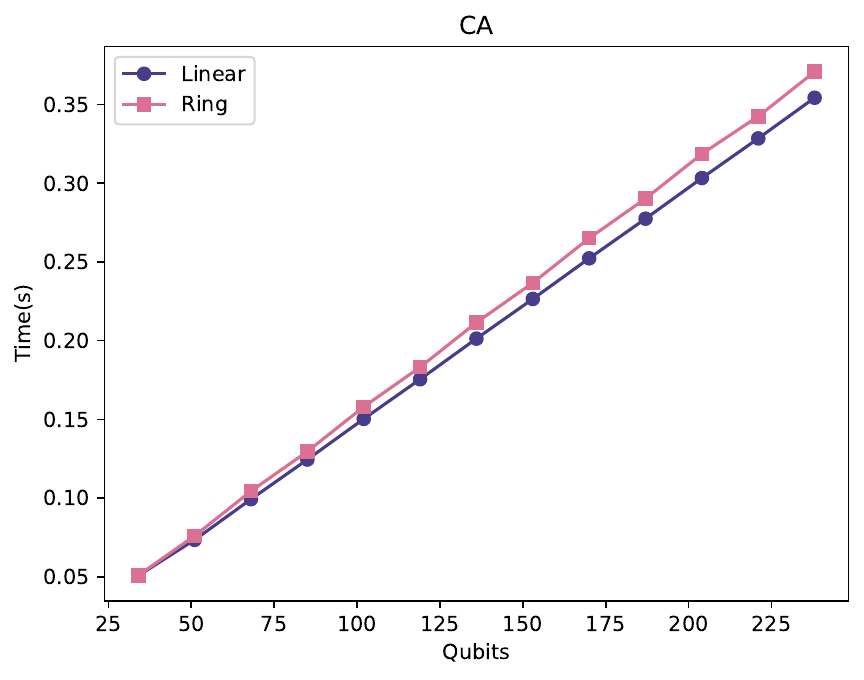}\label{fig:sub1}
        }
    \subfloat[]{
        \includegraphics[width=0.49\columnwidth]{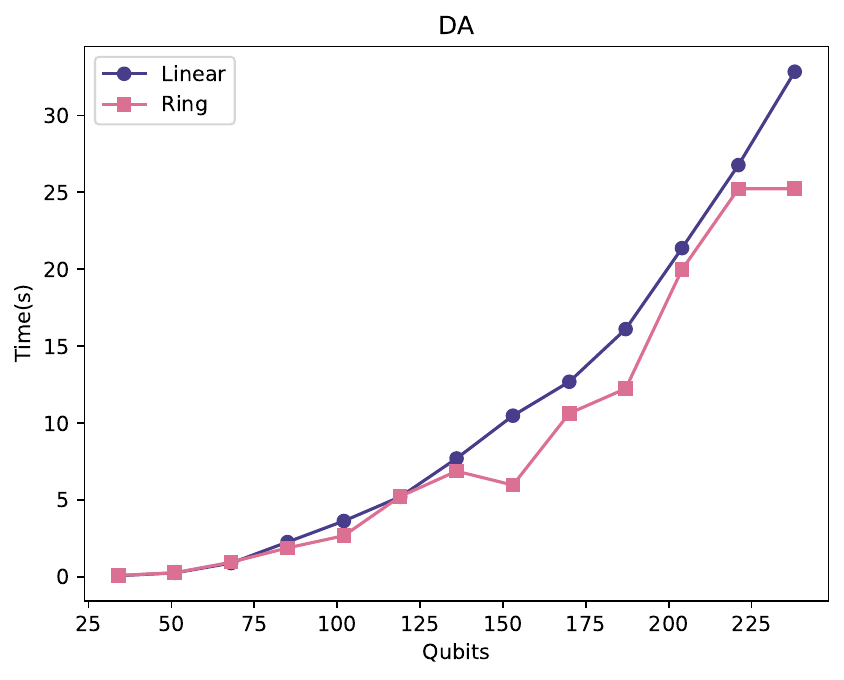}\label{fig:sub2}
    }
    \subfloat[]{
           \includegraphics[width=0.49\columnwidth]{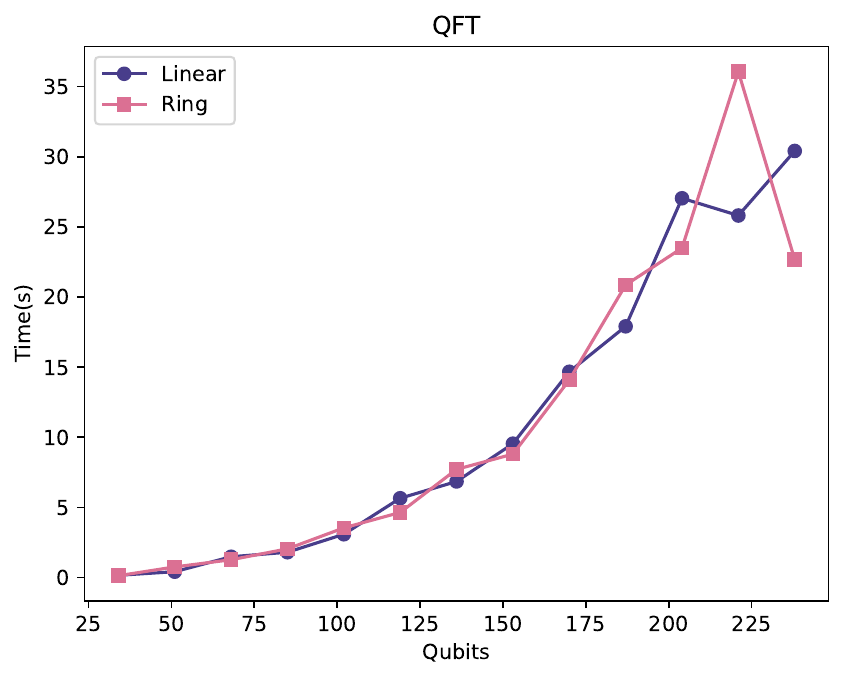}\label{fig:sub5}
    }
     \subfloat[]{
           \includegraphics[width=0.49\columnwidth]{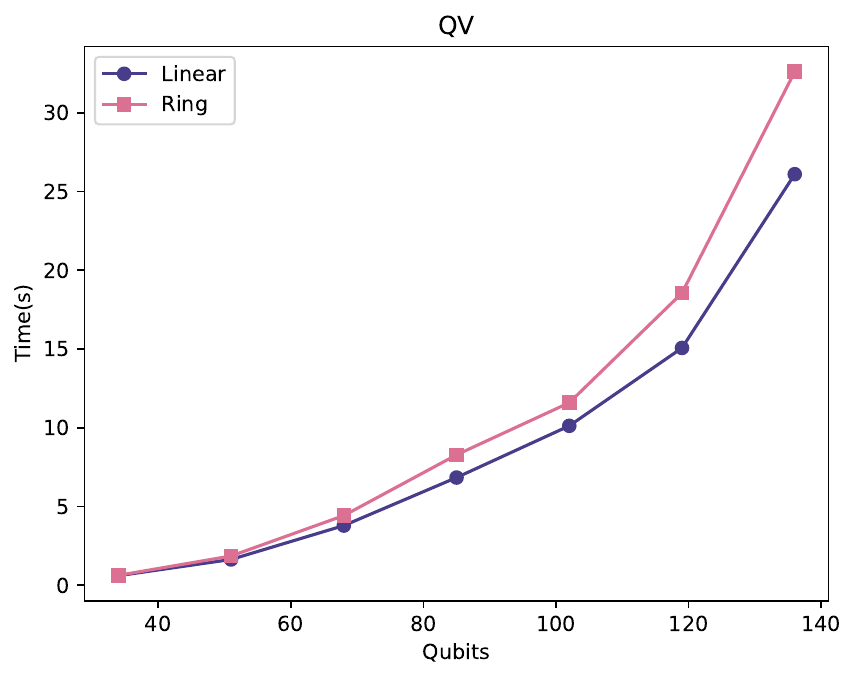}\label{fig:sub5}
    }\\
        \vspace{-12pt}
    \subfloat[]{
        \includegraphics[width=0.49\columnwidth]{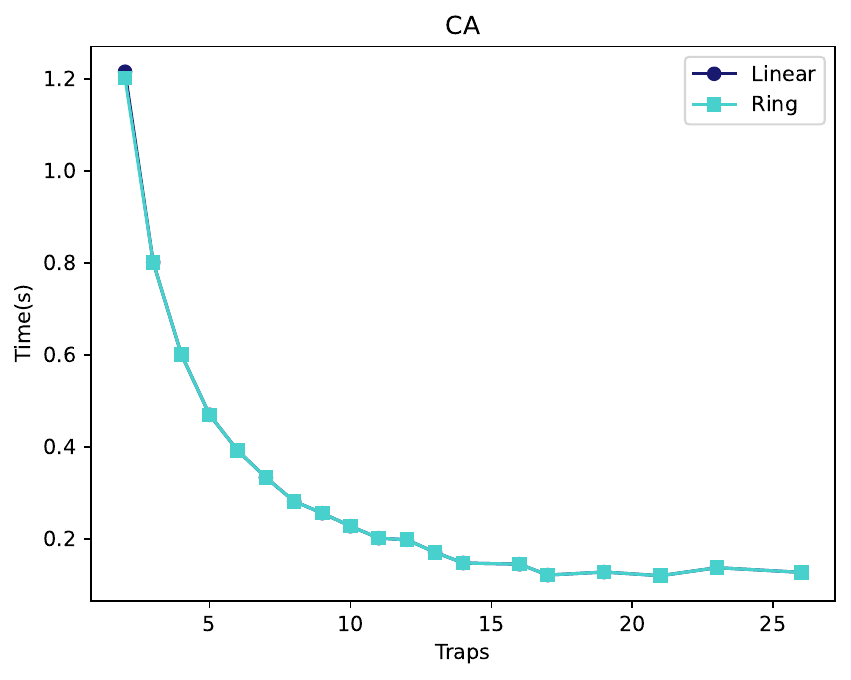}\label{fig:sub1}
        }
    \subfloat[]{
        \includegraphics[width=0.49\columnwidth]{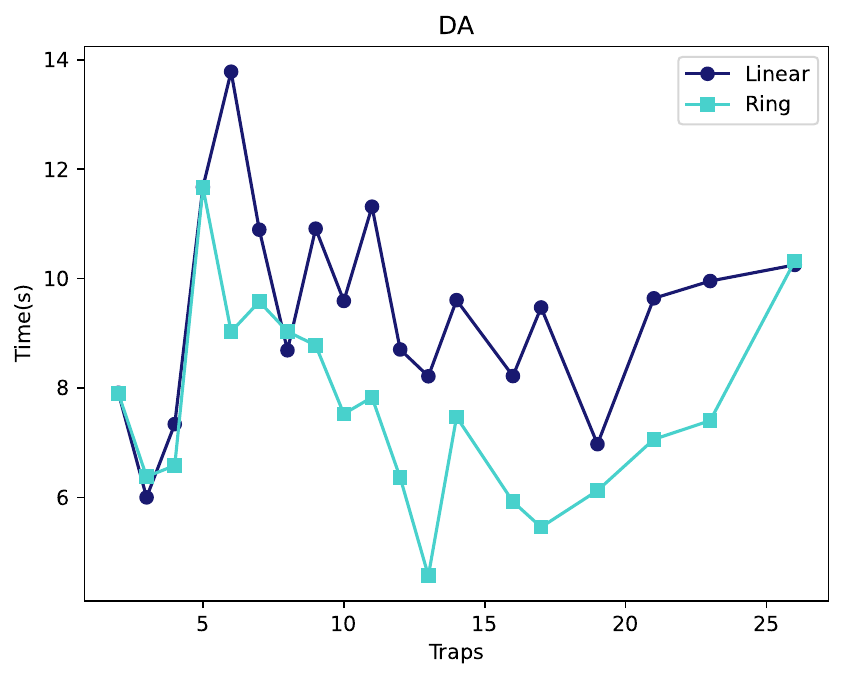}\label{fig:sub2}
    }
    \subfloat[]{
           \includegraphics[width=0.49\columnwidth]{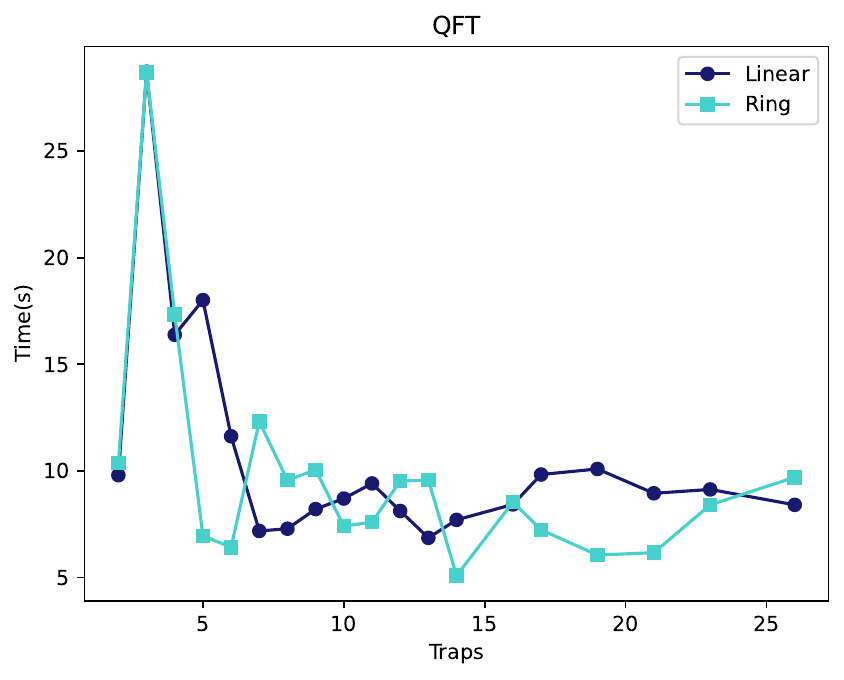}\label{fig:sub5}
    } \subfloat[]{
           \includegraphics[width=0.49\columnwidth]{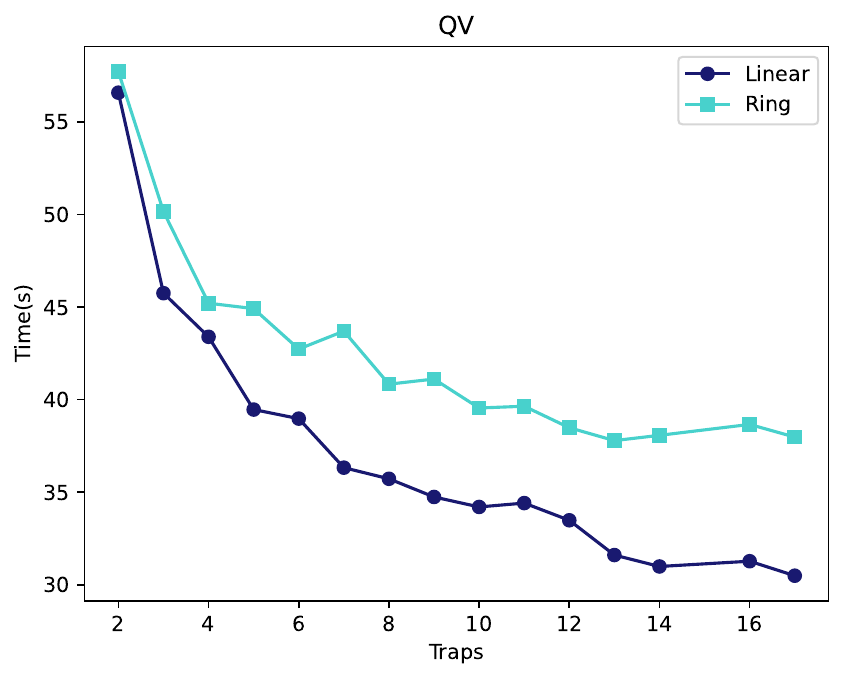}\label{fig:sub5}
    }\\
    \vspace{-7pt}
    \caption{\small{
Scalability analysis for a 1D-linear array and a ring topology (a), (b), (c), (d) strong scaling, and (e), (f), (g), (h) weak scaling.}}
    \label{fig:bench_comp}
\end{figure*}

A scalability analysis was conducted for both topologies, linear and ring, employing two distinct methodologies: strong and weak scaling. This approach enabled the examination of how the various topologies behave as both the number of qubits per trap and the number of traps vary.

Due to the similarities between QAOA and QFT, only results for QFT will be presented. Additionally, given that QV and random circuits incorporate random components, solely the results for QV will be shown for simplicity. For all experiments, an initial excess capacity of 2 was established, allowing for the allocation of additional ions per trap in the subsequent mapping process. The metric utilized across all experiments is the total execution time, which considers all operations such as gates, splits, merges, and shuttles.

The initial experiment conducted is a \textit{strong scaling}, in which the number of physical qubits is increased by gradually adding more traps, each of them having a fixed number of physical qubits. Each trap contains 17 ions, and the number of traps is raised from 2 to 14, reaching a total of 238 physical qubits within the device. However, considering the excess capacity set at the beginning of the execution, benchmarks are chosen to use 210 ions (logical qubits). For QV, being the heaviest quantum algorithm, strong scaling was performed until it reached 140 physical qubits due to the very long simulation time.

Figure~\ref{fig:bench_comp} (top row) shows the results of the strong scaling experiment. Across all benchmarks, a consistent trend is observed: the Cuccaro Adder (CA) exhibits linear behavior, while the remaining benchmarks display exponential growth. As the number of qubits and traps increases, the execution time correspondingly rises, as anticipated, due to the increased number of operations required to execute the algorithms. Additionally, both topologies demonstrate similar trends, with the ring topology slightly outperforming the linear topology for DA, whereas the linear topology exhibits marginally better performance for CA and QV.

The subsequent experiment is a \textit{weak scaling}, in which the total number of physical qubits in the device is kept constant at 180 ions; meanwhile, the number of traps increases, redistributing the physical qubits among the traps accordingly. The benchmarks executed utilize only 128 qubits due to the initial excess capacity allocated for each trap -2 ions. The experiment begins with two traps of 90 ions each. Subsequently, the number of traps is incremented by one each time, while the total number of ions in the device remains unchanged (i.e. the number of traps is increased while decreasing the number of ions per trap). This results in the distribution of ions across all traps until reaching 26 traps, with 6 ions per trap, always keeping 2 ions for the excess capacity, except for the QV, where the experiment was performed until it reached 17 traps due to the long simulation time.

The results are depicted in Figure~\ref {fig:bench_comp} (bottom row). It can be seen that, in half of the cases, the ring topology outperforms the linear one (DA and QFT). For CA, the execution times are similar. This can be attributed most probably to the easy execution of the algorithm as it does not require a lot of complexity in the mapping procedure, thereby resulting in negligible differences between the two topologies.

In the DA and QFT cases, the execution time initially decreases before subsequently increasing again. This phenomenon arises from increased parallelism; specifically, a higher number of traps allows for more concurrent executions. It is important to recall that operations within a single trap must be executed sequentially, thus a higher number of traps leads to enhanced parallelism. However, with more traps, the number of qubits per trap decreases, resulting in a higher frequency of shuttle and SWAP operations, consequently increasing the total execution time. In these cases, the optimal QCCD architecture comprises approximately 14 to 16 traps and between 11 to 13 ions per trap, considering an initial excess capacity of two.

 In the QV results, the execution time decreases when the number of traps increases; this behavior can be attributed to its high level of gate parallelization, as shown in Table~\ref{tab:Benchmarks}. With more traps, more operations can be concurrently executed, leading to a notable reduction in the overall execution time, as occurs in DA and QFT cases. 

It is crucial to note that despite the variations in results, a consistent trend is observed for all benchmarks. This consistency suggests that the device design can be refined to achieve an overall optimization across a wide range of quantum algorithms. Additionally, it should be considered that these results may be influenced by the chosen initial excess capacity and mapping approaches, and there is potential to optimize the outcomes accordingly.

\subsection{Evaluating the impact of the excess capacity}

The excess capacity was analyzed to determine the optimal number of free spaces for each trap at the start of executions for both the 1D-linear array and ring topologies. Similar to previous experiments, the Cuccaro Adder (CA), Draper Adder (DA), Quantum Volume (QV), and Quantum Fourier Transform (QFT) were utilized. In all cases, benchmarks were executed using 64 logical qubits, with the total execution time serving as a metric.

Two distinct approaches were implemented for this experiment. In the first one, the excess capacity was augmented by adding additional ions, from 1 to 10, to a fixed number of traps. There are 5 traps, initially each of them containing 14 ions. The experiment concluded with 24 ions per trap, with 10 ions allocated as excess capacity at the start of the execution. In the second approach, the total number of qubits per trap decreased as the excess capacity increased. Specifically, with a total of 14 ions per trap, the excess capacity was gradually increased to 10, resulting in 4 qubits per trap. Consequently, this approach utilized more traps compared to the first approach, totaling 16 traps. However, at the beginning, only 5 traps were utilized.

\begin{figure*}[!h] 
     \centering
    \subfloat[]{
        \includegraphics[width=0.49\columnwidth]{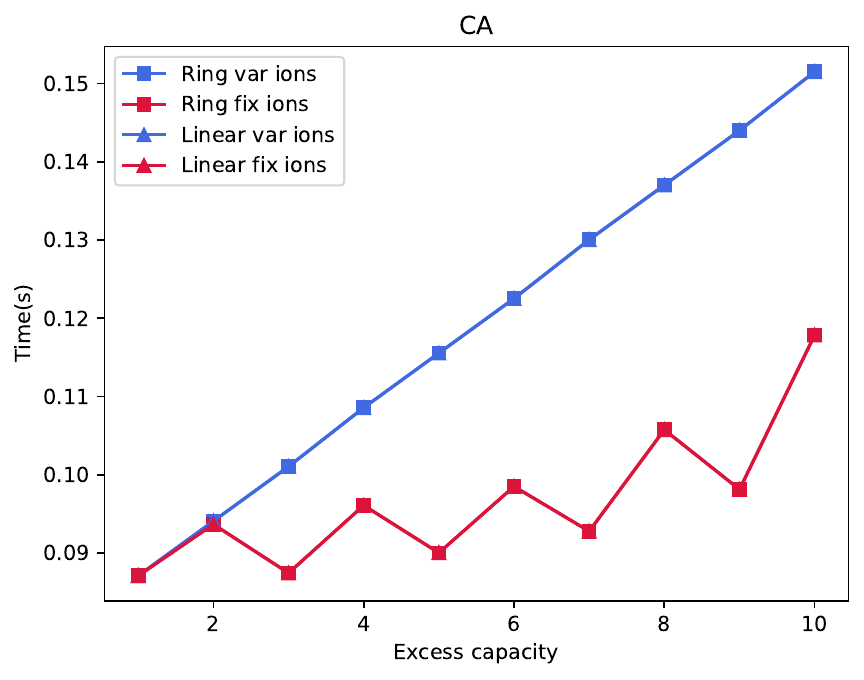}\label{fig:sub1}
        }
    \subfloat[]{
        \includegraphics[width=0.49\columnwidth]{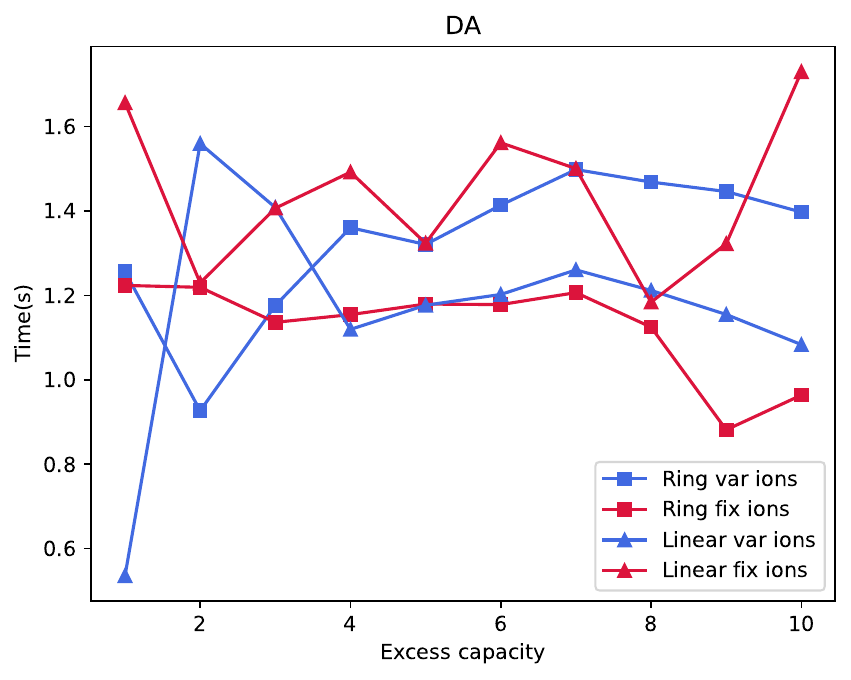}\label{fig:sub2}
    }
    \subfloat[]{
           \includegraphics[width=0.49\columnwidth]{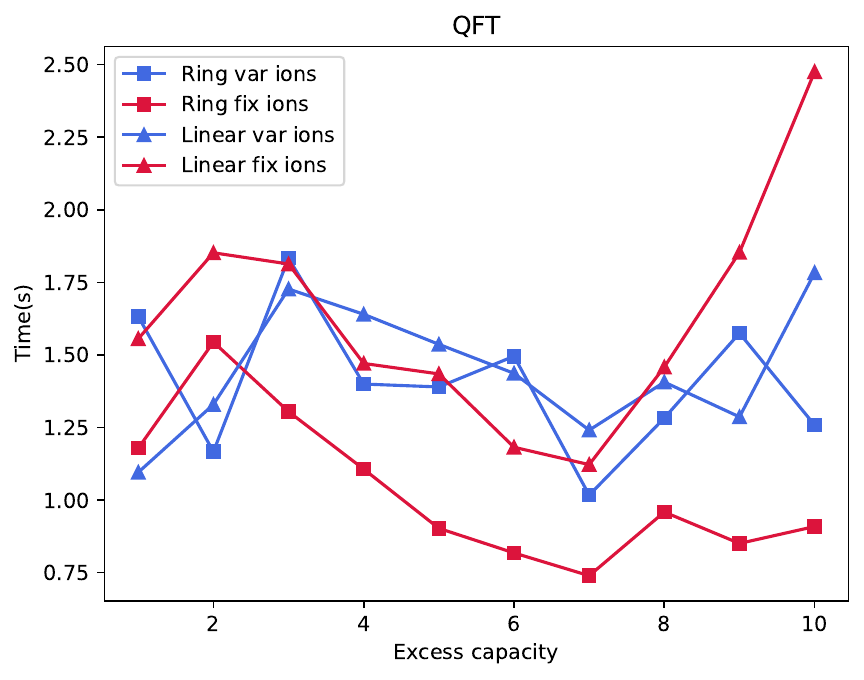}\label{fig:sub5}
    }
     \subfloat[]{
           \includegraphics[width=0.49\columnwidth]{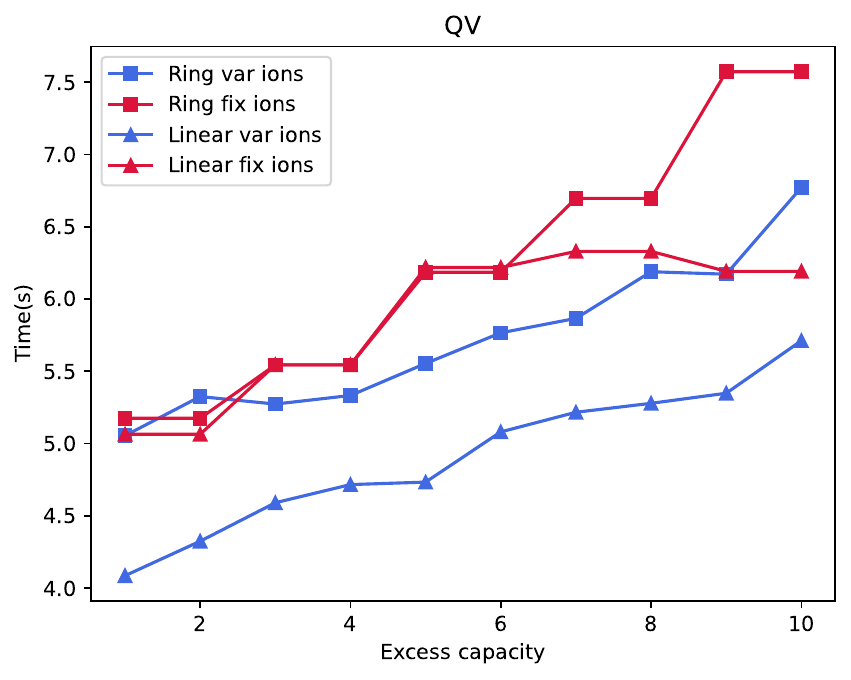}\label{fig:sub5}
    } \\
        \vspace{-12pt}
    \subfloat[]{
        \includegraphics[width=0.49\columnwidth]{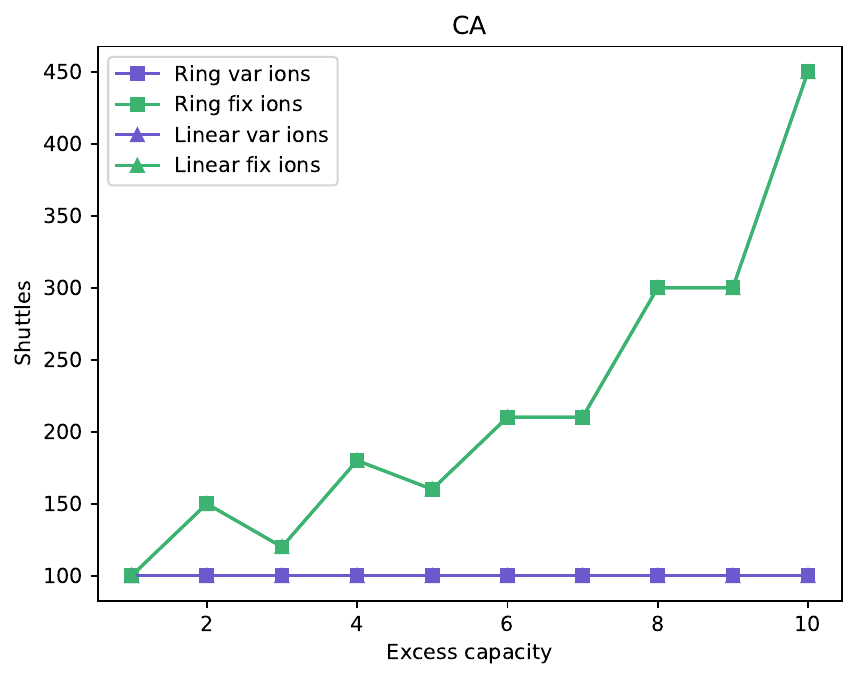}\label{fig:sub1}
        }
    \subfloat[]{
        \includegraphics[width=0.49\columnwidth]{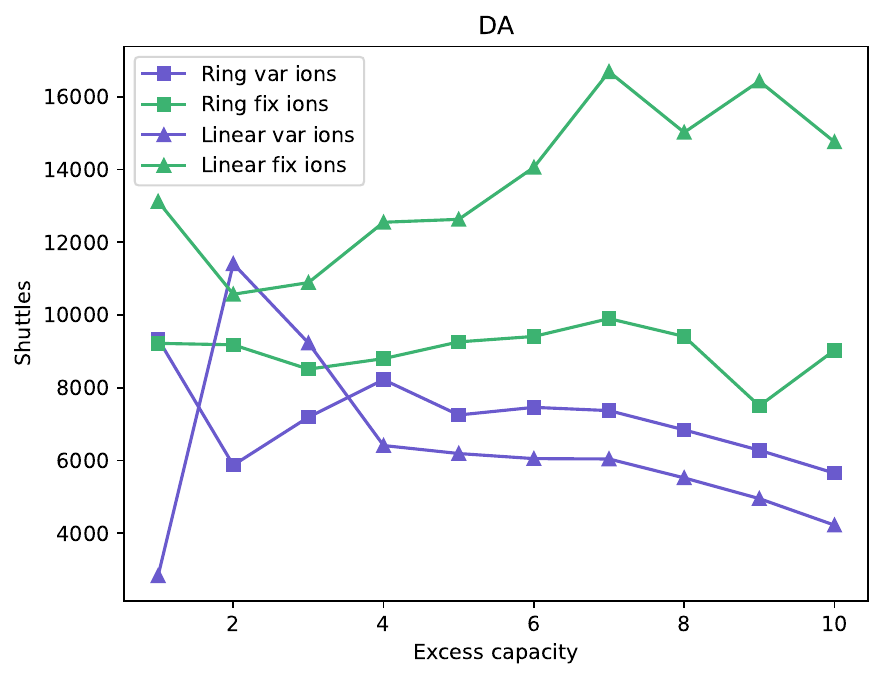}\label{fig:sub2}
    }
    \subfloat[]{
           \includegraphics[width=0.49\columnwidth]{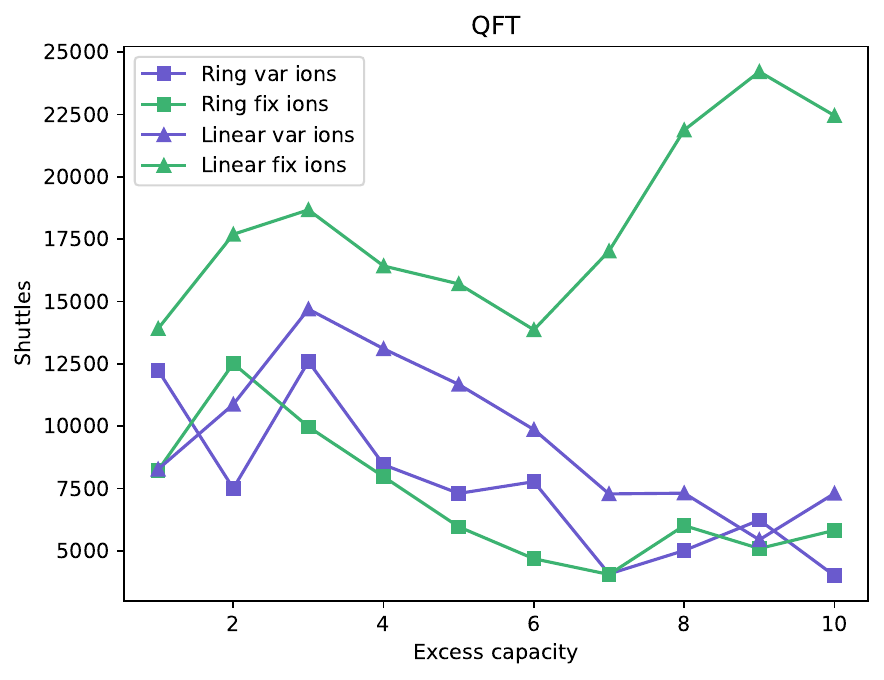}\label{fig:sub5}
    } \subfloat[]{
           \includegraphics[width=0.49\columnwidth]{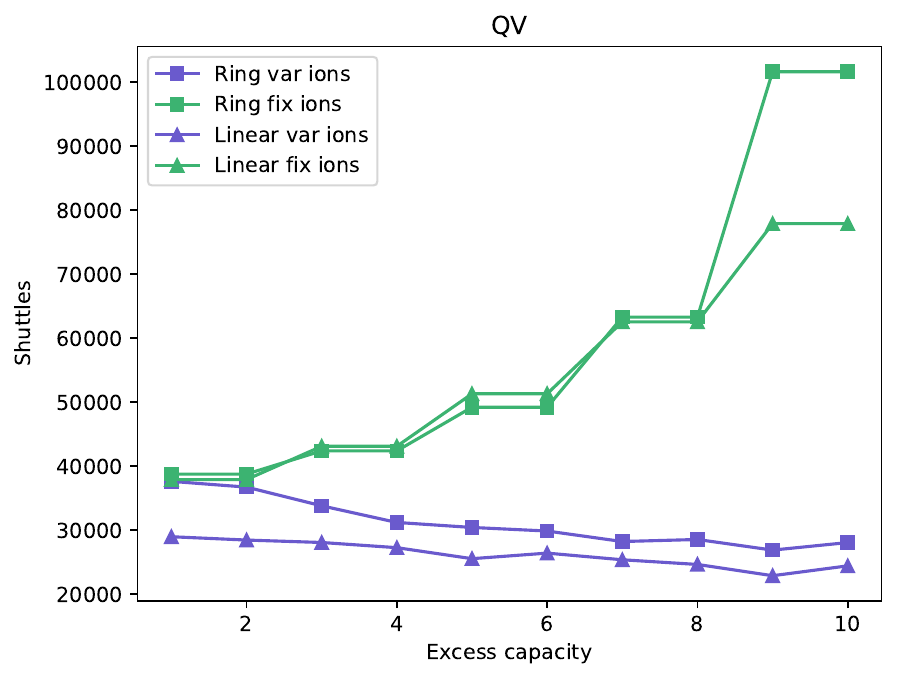}\label{fig:sub5}
    }\\
    \vspace{-7pt}
    \caption{\small{Excess capacity analysis for the 1D-linear array and the ring. Two experiments are shown: (i) a fixed number of ions per trap (fix ions) and (ii) increasing ions per trap with the excess capacity (var ions). First row shows execution time and second row number of shuttles. }}
    \label{fig:EC_comp}
\end{figure*}

The results for both experiments are shown in Figure~\ref{fig:EC_comp}. In the case of CA, for both topologies, when the number of ions is fixed in a trap and the excess capacity varies, a trend is observed where the results improve when an even number of excess capacities is allocated. This is caused by the STA algorithm and the CA structure: a qubit from the ratio list is placed when one spot is left in a trap, and the excess capacity is even; in contrast, a pair from the time interaction list is placed when a trap is becoming full, and the excess capacity is odd, leading to the observed behavior. As the excess capacity increases alongside the ions per trap, a linear trend is observed. In this case, the number of shuttles and SWAPs is equal to zero, as qubits do not need to move traps, as shown in Figure~\ref{fig:EC_comp} (only the results for shuttles are shown due to similar trends). However, the time required to execute these operations increases due to the higher number of ions per trap. Specifically, a greater execution time is needed to perform gates as the number of qubits per trap increases.

For DA, the ring topology with a fixed number of ions per trap emerges as the most optimal approach, whereas the linear topology with a fixed number of ions per trap ranks as the least favorable. In general, a linear topology with excess capacity equal 1 and a ring topology with an excess capacity of 2 and 14 qubits per trap appears to be best performing configuration. Conversely, although the number of SWAPs and shuttles decreases as the number of ions increases with respect to the excess capacity, similar to the CA scenario, the execution time increases due to the greater number of ions within a trap, rendering it a less optimal approach. An intriguing observation is that a total fixed number of ions per trap for both topologies yields opposite results. This can be attributed to the superior connectivity of the ring topology compared to the linear one, resulting in fewer shuttles and SWAPs movements without increasing the execution time.

A similar trend to the results obtained for DA is observed in QFT. The ring topology with a fixed number of ions per trap emerges as the most efficient approach, while the linear topology with an increasing number of ions per trap proves to be less optimal. However, a notable difference between QFT and DA is noted in the behavior regarding excess capacity in the ring topology. Initially, increasing the excess capacity leads to improved efficiency until reaching a maximum at an excess capacity of 7, after which it diminishes. This phenomenon can be attributed to QFT's higher degree of gate parallelization compared to DA, as shown in Table~\ref{tab:Benchmarks}. As excess capacity increases, qubits are more evenly distributed across traps, enabling more parallel executions. However, beyond a certain point, execution time starts to increase. This is due either to the high number of ions in a single trap (limited parallelism and increase in gate execution time) or to too few qubits placed in each trap due to the high excess capacity (increasing the number of SWAPs and shuttles). It is important to highlight that the routing process does not prioritize parallelization when relocating ions. Consequently, distributing ions evenly across traps naturally enhances parallelization, a factor not explicitly considered in the routing algorithm.
Lastly, in the QV benchmark, a different trend is observed compared to all other benchmarks: the most optimal approach is to increase the excess capacity with the number of ions per trap. Through the routing process, ions are stored within the fewest number of traps possible, thereby reducing the total number of shuttling and SWAP operations. While this approach does not yield favorable results for other benchmarks, it proves effective for QV by significantly reducing the overall number of operations. Note that in this case is more efficient to keep all qubits in fewer traps as QV shows a high degree of qubit interactions. Through experimentation with excess capacity, it becomes evident that the optimal initial free spaces depend on the quantum algorithm to be executed. Additionally, it is observed that, overall, the ring topology tends to show better results compared to the linear one. However, careful consideration must be given when adding extra ions per trap, as this can significantly diminish the quality of executions, even if those ions are not utilized during the execution of the quantum algorithm on the device.
\section{Conclusion and future work}
This paper introduced a new qubit allocation strategy for traped-ion QCCD architectures, the "Spatio-Temporal Aware Qubit Allocation" (STA) algorithm, which demonstrates an enhancement of up to 50\% in execution time compared to previous methods. Additionally, we conducted a scalability analysis (weak and strong scaling) for two distinct topologies: a 1D-linear array and a ring topology. Results indicate that as the number of qubits and traps increases, so does the execution time. However, ion distribution across an optimal number of traps enhanced parallelism, thereby reducing execution time. Our findings suggest that current ion trap qubit routing techniques could be further optimized for ring topologies. Furthermore, the consistent trends observed across all algorithms in the scalability analyses indicate that architectural designs and mapping procedures can be optimized to enhance the overall performance across a diverse range of quantum algorithms. Lastly, the initial excess capacity was explored, showing a very different behavior for each quantum algorithm.

Further research should be done to improve the qubit routing procedure in modular QCCD devices for both topologies, linear and ring, with a particular focus on optimizing the excess capacity.  In addition, there is potential to refine mapping techniques specifically tailored for ring topologies, capitalizing on their superior connectivity relative to 1D linear arrays.

\section*{Acknowledgments}
The authors acknowledge financial support from the European Union’s Horizon Europe research and innovation program through the project Quantum Internet Alliance under grant agreement No. 101102140. CGA also acknowledges support from the Spanish Ministry of Science, Innovation and Universities through the Beatriz Galindo program 2020 (BG20-00023) and the European ERDF under grant PID2021-123627OB-C51.



\bibliographystyle{IEEEtran}
%
\bibliography{bib/main}

\end{document}